\documentstyle[prb,aps]{revtex}
\draft

\begin{document}
\author{I.I. Mazin and Ronald E. Cohen}
\title{Insulator-metal transition in solid hydrogen: Implication of electronic
structure calculations for recent experiments. }
\address{Geophysical Laboratory and Center for High-Pressure Research\\
Carnegie Institution of Washington,
5251 Broad Branch Rd. N.W., Washington D.C., 20015.}
\date{June 6, 1995 }

\twocolumn[\maketitle

\begin{abstract}
Electronic structure calculations for compressed molecular hydrogen are
performed to provide more insight into the diversity of phenomena recently
observed experimentally. We perform full-potential LAPW calculations and
analyze them in terms of a molecular tight-binding model. We show that $%
\sigma _g$ and $\sigma _u$ bands overlap occurs at rather low pressure,
while an insulating state persists at specific orientations up to 200 GPa,
in accord with previous work, and is due to opening of hybridization gaps at
the Fermi level. We also present calculations of electronic properties such
as plasma frequencies and electron-vibron coupling.
\end{abstract}


]

\narrowtext
Experimental data on the behavior of hydrogen at megabar pressures is
rapidly emerging as a result of advances in in diamond-anvil cell techique.
Certain results are clearly established and understood, while a number of
questions remain, most of which concern the implication of spectroscopic
results for the electronic structure. One issue, in particular, is band
overlap metallization. It is well established that at room temperature and
below hydrogen under high pressure stays insulating up to at least 150 GPa%
\cite{RMP} (which is about tenfold compression, $r_s\approx 1.4$). LDA
calculations\cite{Ash,K,Japs,Bar} give a metallization pressure that
strongly depends on the assumed orientation of hydrogen molecules (reported
numbers are from 40 to 150 GPa), the
strongest metallization tendency being in the structure where all hydrogen
molecules are parallel. The reason for such a strong orientational
dependence is not completely understood.

A very basic argument for band-overlap metallization of hydrogen is that the
insulator-metal transition occurs when the bandwidths of the 1$\sigma _g$
and $1\sigma _u$ bands originating from the corresponding molecular orbitals
become larger than the splitting of the two molecular levels. One can easily
estimate when this should happen using, for instance, Slater's variational
LCAO\cite{slater}, and assuming the perpendicular orientation (which minimizes
the overlap) of neighboring
molecules\cite{calc}. The answer is $r_s\approx 1.45.$ More accurate
numerical calculations, starting from Friedli and Ashcroft,\cite{Ash},
have yielded similar numbers. However, it turns out
that even LDA calculations (see, e.g.,
 Refs.\onlinecite{Ash,K,Japs,Bar}) render an
insulating ground state {\it in the orientation which minimizes the
 total energy}.
In the local density approximation (LDA), where molecular orbitals are too
diffuse, this sort of metallization occurs at even larger $r_s$.

A detailed
microscopic analysis of compressed hydrogen, even in LDA and for clamped
nuclei, has not been yet presented. To provide further insight into the
available experimental data, we performed full-potential LAPW calculations
for hydrogen for different orientations. By mapping the calculated band
structure onto the two molecular orbitals nearest-neighbor tight binding
model we show how this band structure is formed and why the minimum total
energy orientation appears insulating.

Most previous calculations\cite{K,Japs,Bar} used a plane wave expansion for
the wave functions. This is a good basis set for the interstitial region;
however, close to nuclei it converges slowly. For a standard cut-off of 50 Ry
the Fourier expansion of hydrogen 1$s$ wave function converges at a nucleus
only within 14\%. We used the Linearized
 Augmented Plane Wave method with an MT
radius of 0.69 bohr; for $r\geq 0.69$ the error in the aforementioned 1$s$
Fourier expansion is about 50 times smaller than for $r=0$. We found the
total energy to be convergent better than 1 mRy, with the cut-off of 44 Ry.
We find that the calculations converge relatively
slowly with the number of {\bf k}-points in the Brillouin zone; we had to
use a special {\bf k}-point mesh 12$\times 8\times 8$ corresponding to 96
points in the irreducible edge for the 4-molecule orthorombic structure
described below. Most calculations reported here were
performed for $r_s=1.435$ (lattice parameter 3.3 bohr, molecular volume
2.209 cm$^3$/mol, $\rho /\rho_0=10.5$).
The calculated pressure was about 150 GPa at this
compression. We also performed some calculations at higher compressions in
order to detect the LDA metallization.

The crystal structure of high-pressure phases II and III has not been
determined\cite{RMP}. It is generally believed that centers of molecules
form an hcp lattice, as they are known to do to at least 50GPa. At least
four distinct optical (Raman and infrared) vibron modes are observed in
phase II in hydrogen and deuterium\cite{RMP,II}, indicating an
orientational ordering of at least four molecules per cell. Phase III has an
orientational order different from that of phase II\cite{we}, with one
strong Raman and one extremely strong infrared vibron\cite{RMP}. Both
vibrons show substantial softening, typically by 200 cm$^{-1}$ between 200
and 100 K, when passing into phase III. The structure
considered the most relevant for the high-pressure phases of hydrogen is the
Pca2$_1$ structure which is derived from the hcp-c (in which all molecules
are aligned parallel to $c$, space group P6$_3$/mmc) by doubling the unit
cell and making the four molecules in the unit cell to be tilted from $c$ by
an angle $\theta $ and rotated by $\pm \varphi ,\pi \pm \varphi $. This
structure is known to minimize the electrostatic energy of classical
quadrupoles on an hcp lattice, at $\theta \approx 55^{\circ }$ and $\varphi
\approx 43^{\circ }$. Notably, even at $\rho /\rho_0\approx 13$ ($r_s\approx
1.35)$
the equilibrium orientation, calculated from total energy methods, is
similar to this\cite{Japs}, although quadrupolar interaction is is not the
dominant anisotropic interaction at this compression. Our computations,
described below, give $\theta \approx 62^{\circ }$ and $\varphi \approx
50^{\circ }$ at $r_s=1.435$.

The assumed structure has five degrees of freedom: $c/a$ and $b/a$, two
orientational angles, and the H$_2$ bond length. We have fixed $b/a=\sqrt{3}$%
, as in hcp, and optimized the structure with respect to four remaining
parameters. We find that $c/a$ is not sensitive to orientation: for the
high-symmetry P6$_3$/mmc structure, i.e. for $\theta =0$, we find $c/a=1.59$%
, while for ($\theta ,\varphi )=(62^{\circ }$, $50^{\circ })$ orientation it
is 1.62. The optimal bond length changes more: for the same $c/a$ (1.59) the
bond length $d$ in P6$_3$/mmc structure is 1.42 bohr, while for ($62^{\circ
} $, $50^{\circ })$ it is 1.48 bohr. Most of our calculations below use $%
c/a=1.6$ and $d=1.5$ bohr.

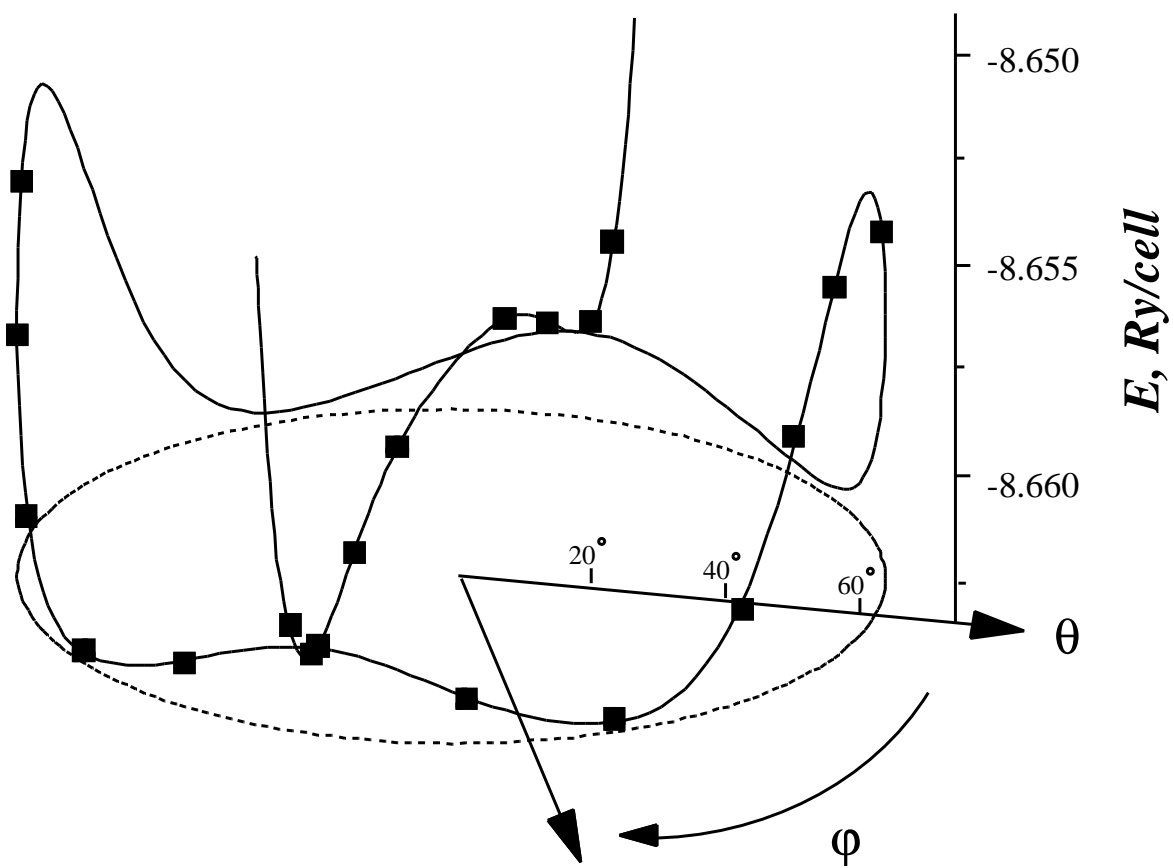
\begin{figure}
\caption{Orientational dependence of the total energy of Pca2$_1$
hydrogen. Radial coordinate,$\theta $, is the polar angle and angular
coordinate, $\varphi $, is the azimutal angle. Lines are guides for eye.}
\label{rot}\end{figure}

Fig.\ref{rot} shows the orientational dependence
of the total energy. Note that the potential barrier for librations around ($%
62^{\circ }$, $90^{\circ })$ is rather low, about 280 K. Of course, the
calculated energy profile corresponds to coherent rotations. Actual behavior
of the system depends on how the energy profile differs when a single
molecule is rotating in the field of its neighbors.
 For a single molecule librating in the potential shown
in Fig.\ref{rot}, we can solve the Schr\"{o}dinger equation for the
anharmonic oscillator numerically and
find that even at zero temperature the molecule would be freely librating
around the average orientation ($62^{\circ }$, $90^{\circ })$ with the
amplitude of $\approx $40$^{\circ }$ (both for hydrogen and deuterium), thus
making a less orientationally ordered phase. If we want to characterize the
degree of orientational order, we should introduce an orientational order
parameter, which is usually written as a zero-trace tensor $Q_{\alpha \beta
}=(3\langle n_\alpha n_\beta \rangle -\delta _{\alpha \beta })/2$, where
{\bf n} is the unit vector of the molecule's direction. Principal values of
this tensor for axially symmetric librations are \{$Q,-Q/2,-Q/2\},$ and $Q=1$
for full ordering. Librations described above correspond to $Q_{\alpha
a}\approx \{0.5,0,-0.5\},$ while free rotation around $c$-axis with $\theta
=62^{\circ }$ gives $Q_{\alpha a}=\{-0.17,0.085,0.085\}.$

The first two azimutal librational eigenstates are about 30 cm$^{-1}$(7.5 cm$%
^{-1}$) and 170 cm$^{-1}(100$ cm$^{-1})$ above the saddle point at ($%
62^{\circ }$, $90^{\circ })$ for hydrogen (deuterium). Since ($62^{\circ }$,
$90^{\circ })$ is a saddle point, azimutal and polar librations are quite
different: Azimutal librations, as described above, are strongly anharmonic,
while polar vibrations, to the contrary, can be treated in the harmonic
approximation. Their amplitude reaches $\simeq 40^{\circ }$ at $T\simeq 400$%
K. The calculated frequency of the polar libron is about 350 cm$^{-1}$.

One of the most intriguing issue is whether or not the equilibrium geometry
is governed by electrostatic forces. The calculated ground state ($62^{\circ
}$, $50^{\circ })$ is very close to the equilibrium orientation ($55^{\circ }
$, $43.5^{\circ })$ for the quadrupolar hcp lattice (see, also, Ref.\onlinecite
{Japs}). This orientation ($62^{\circ }$, $50^{\circ })$ maximizes the
insulating gap (actually even small deviations from this orientation render a
metallic state). Thus, the band structure energy also favors this orientation.
Besides, at the tenfold compression the
actual electrostatic interaction differs
substantially from the ideal quadrupolar one. These arguments cast doubt
about quadrupolar interaction being the main ordering factor. More insight
can be gained from comparing calculations with different number of ${\bf k}$%
-points. We have performed calculations with a sparser mesh (12 points) and
found that the total energy is minimal at ($20^{\circ }$, $90^{\circ })$. 12
{\bf k}-points for an 8-atoms unit cell is a sufficiently large number to
have the charge density converged. Indeed, we found no substantial
difference between the orientational dependence of the electrostatic energy
in the both sets of calculations. To the contrary, the one-electron energy
orientational dependence is quite different in the two cases (with 12
points, it is not converged yet). Thus, we conclude that the main factor
controlling ithe orientational ordering at about tenfold compression is
one-electron (band) energies.

Since the driving force for orientational ordering is the band energy, it is
useful to compare the electronic bands for several different orientation.
Fig.\ref{bands}a shows the bands
(full lines) for the two-molecules-per-cell P6$_3$/mmc hcp structure.
 To understand the nature of the
relevant bands, we compare the self-consistent band structure with that from
two-molecular-orbital tight-binding (TB) model (dashed lines). The TB
parameters were $t=3$ eV, $E_{\sigma _u}-E_{\sigma _g}=16.8$ eV at $r_s=1.435
$. Only 12 nearest neighbors were taken into account. The lowest states at $%
\Gamma $ and X are pure $\sigma _g$ [actually, $\sigma _g(1)\pm \sigma _g(2)$%
]. However, the next level at $\Gamma $ is already the bottom of the $\sigma
_u$ band. Thus the overlap of the $\sigma _g$ and $\sigma _u$ bands is very
large, of the order of one Rydberg. Note also the third  band at $\Gamma $,
which is the bottom of the 2$\sigma _g$ band.

Since the $\sigma _g-\sigma _u$ band overlap occurs at much lower
compression ($\rho /\rho _0\sim 7)$, H$_2$ would be metallic if not for the
gaps that open at the Fermi level due to the orientational tilting. Fig.\ref
{bands}b shows bands for a less symmetric ($62^{\circ }$, $90^{\circ })$
structure (dashed lines), compared to the folded down hcp bands (dotted
lines). At several points where two downfolded bands cross in the hcp
structure, the degeneracy is lifted now because of lower symmetry. This band
structure is still metallic; the crossings occur
at different energies. Nevertheless, the density of states at the Fermi
level and correspondingly the one-electron energy is decreased.

If we now move from the ($62^{\circ }$, $90^{\circ })$ structure further to
the equilibrium orientation ($62^{\circ }$, $50^{\circ }),$ (full lines in
Fig. \ref{bands}b) we observe that
local minima in the conduction band, and maxima of the valence band (which
emerged from the band crossings in the P6$_3$/mmc structure) are now aligned
at, respectively, 18 eV and 17 eV, thus forming a considerable gap. Rotating
the H$_2$ molecule influences the positions of these maxima and minima in a
complicated way; alignment of the direct gaps such that an indirect gap
appears is characteristic for the equilibrium orientation. There is an
analogy with the Peierls transition in metals with Fermi surface nesting: if
a superstructure exists in which the nesting bands become degenerate,
lowering of symmetry removes the degeneracy, decreases the band energy and
induces a structural phase transition. An important difference is that there
is {\it no} nesting, and therefore the regular susceptibility would diverge
in the high-symmetry phase, as in a Peierls transition. Instead, there is an
instability with respect to final rotation of the molecules.
Correspondingly, the known problem of the one-dimensional Peierls
transition, namely that the divergence in the non-interacting susceptibility
may cancel out in the full susceptibility\cite{yossi}, does not apply.

Another way to rationalize this result is to use Andersen's force theorem%
\cite{oka}, which states that the first-order total energy differences in
LDA can be calculated as a sum of the electrostatic energy differences and
band energy differences, provided that charge density is not recalculated to
self consistency in the perturbed configuration (for instance, if one would
ascribe a sphere around each molecule and rotate it rigidly without changing
the charge distribution inside the sphere). While it was not done in our
calculations (and is technically not possible in LAPW method), the changes
of molecular charge density upon rotation are small.

We have also performed calculations for higher compression, but without
orientational optimization. The indirect gap closes after further
compression by 6.5\% ($P\approx 185$ GPa). It is well known\cite{IEG} that
the LDA considerably underestimates semiconductor gaps. Therefore, not only
should hydrogen stay insulating in the pressure range of 150 GPa, but the
energy of the gap is likely to be rather large. Of course, dynamical
disorder smears the gap, thus one can expect gradual increase of the
low-energy absorption with temperature and pressure in a non-Drude manner,
with eventual full metallization at pressures larger than at least 185 GPa.
Optical experiments do not reveal metallic, Drude-like absorption up to $%
\sim $250 GPa\cite{Han91,Eggert91}; In the phase III which exists above 150
GPa, strong activity of the infrared vibron with frequency $\nu \approx
4\times 10^3$ cm$^{-1}$ (in deuterium $\approx 3\times 10^3$ cm$^{-1})$ is
well documented\cite{RMP}. This puts an upper limit to the plasma frequency:
if hydrogen is a semi-metal at this pressure, then $\omega _{pl}\lesssim $
0.4 eV.

The main vibron frequency softens discontinuously at the transition to the
phase III, which in some papers was attributed to the onset of metallic
screening\cite{Bar}, and keeps softening with cooling (up to $\simeq 200$ cm$%
^{-1})$\cite{we}. Current calculations provide several strong arguments
against this point of view. First, as also noticed in previous calculations%
\cite{K,Japs,Bar}, an arbitrary orientation results in a smaller gap or in
the closure of the gap, compared to the optimal orientation. Thus one can
expect that at elevated temperature when the molecules do not assume any
specific orientation, metallization occurs at lower pressure, and upon
heating, not cooling. This would result in hardening of the vibrons with
cooling, and not softening,  and would also demand that phase boundary line
would shift to lower temperature with pressure, opposite to the observed
trend\cite{RMP,we}.

It is also instructive to estimate the possible effect of metallic screening
on vibron frequencies. To do that, we can calculate coupling of the main
vibron with conducting electrons,  $\lambda =-\Delta \omega /\omega =\sum_{n%
{\bf k}}(\partial \epsilon _{n{\bf k}}/\partial R)^2/2M\omega ^2\delta
(\epsilon _{n{\bf k}}-E_F),$ where $\Delta \omega $ is the change of the
vibron frequency $\omega $ due to metallic screening, $R$ is the H-H bond
length, $M$ is the proton mass and the deformation potentials $(\partial
\epsilon _{nk}/\partial R)$ are the derivatives of one-electron energies
with respect to the bond length\cite{COR}. We have done such calculations
for the ($\varphi ,0^{\circ })$ and for ($65^{\circ },90^{\circ })$
orientations, using 96 nonequivalent first-principle {\bf k}-points
interpolated onto the denser mesh of 490 nonequivalent points, using a
Fourier interpolation \cite{Pickett}. By shifting $E_F$ in the valence band
we imitate the effect of different degrees of metallization without doing
calculations at the different pressures. Simultaneously we can control
the degree of metallization by calculating the plasma frequency $\omega
_{pl}^2=(8\pi e^2/\hbar ^2m_eV_{cell})\sum_{n{\bf k}}(\partial \epsilon _{n%
{\bf k}}/\partial {\bf k})^2\delta (\epsilon _{n{\bf k}}-E_F)$ and comparing
it with the above-mentioned upper bound. The results are shown on Fig.\ref
{lambda}. One can see that the effect of possible metallization on the
phonon frequencies is much too small to explain the softening.

To summarize, we presented detailed full-potential LAPW calculations for
clamped-nuclei molecular hydrogen in the Pca2$_1$ crystal structure for
pressures of 150 GPa and above. The structure was optimized with respect to $%
c/a$, H-H bond length, and molecular orientations. We found that orientation
that minimizes total energy is close to the equilibrium orientation of
classical quadrupoles,although the main driving force for orientational
ordering is one-electron (band) energy. Molecular hydrogen remains
insulating well above the compression at which $\sigma _g-\sigma _u$ bands
start to overlap, because at the equilibrium orientation $\sigma _g-\sigma
_u $ band crossings occur close to Fermi energy, so that the hybridizarion
open a dielectric gap. On the other hand, dynamical disorder due to
zero-point motion will transform the dielectric gap into a pseudogap. We
also have calculated electron-vibron coupling constant and found that in
cannot account for the 5\% vibron softeneing in Phase III, which must be
related primarily with new orienational ordering in this phase.

We thank A. Goncharov, M. Li, and H.-K. Mao
for discussions of the high-pressure experiments on hydrogen, and
particularly R. Hemley for numerous useful comments. Computations were
performed on the Cray Y-MP at the NCSA and on the Cray-90 at Pittsburg
Supercomputer Center.

\begin{figure}
\caption{ (a) LAPW energy bands for the hcp structure,
P6$_3$/mmc, compared with a (dash lines) two-molecular- orbitals
tight-binding
model, (b) LAPW energy bands for Pca2$_1$ structure in ($0^{\circ },\varphi$)
orientation (downfolded P6$_3$/mmc structure, dotted lines),
($62^{\circ },90^{\circ }$) (dash lines)
and ($62^{\circ },50^{\circ }$) orientations.}
 \label{bands}
\end{figure}
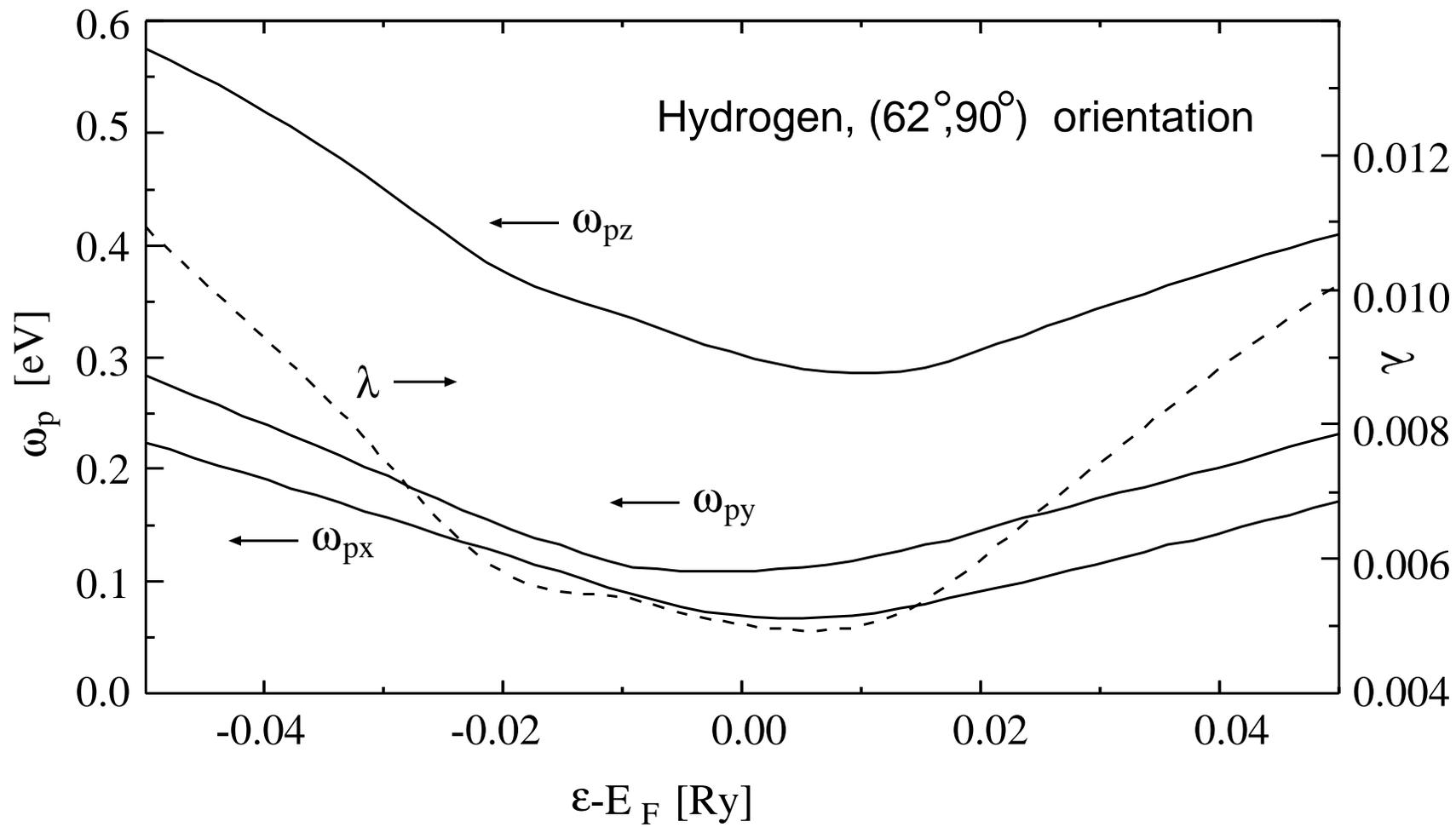
\begin{figure}
\caption{Plasma frequencies and electron-vibron coupling
constant for ($62^{\circ },90^{\circ }$) orientation. Experimental upper
limit for $\hbar \omega _p,\ \approx 0.4$ eV, is also shown.}\label{lambda}
\end{figure}


\begin{references}
\bibitem{calc}  Slater's variational LCAO method for hydrogen approximates 1$%
s$ orbital of a hydrogen atom in a molecule as $e^{\alpha r}$, where $\alpha
$ is a variational parameter ($\alpha \approx 1.31$ for interatomic distance
of 1.5 $a_B$). Overlap integral for nearest-neighbor atoms of different
molecules can be easily calculated and yield the number in the text.
\end{references}
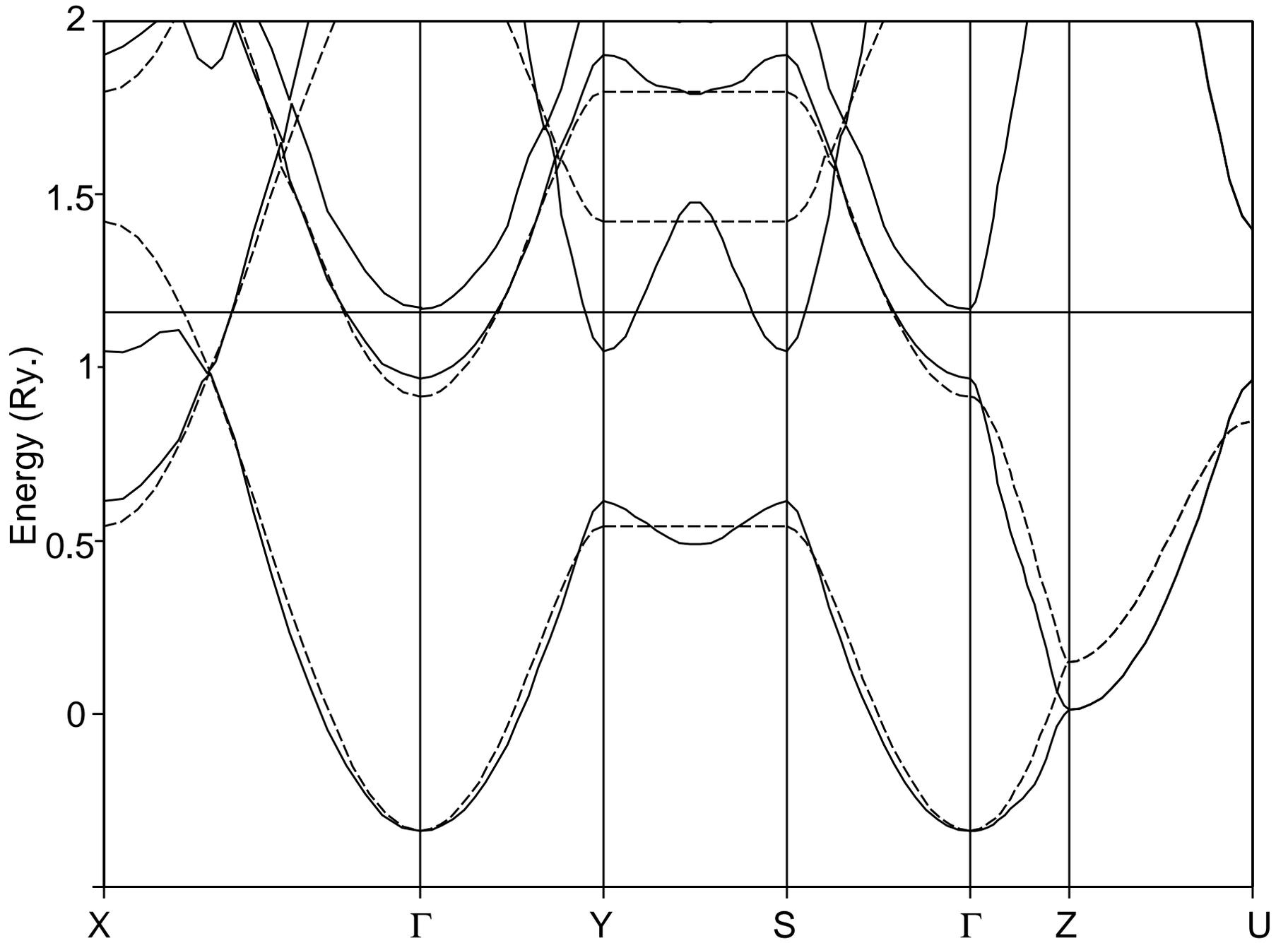
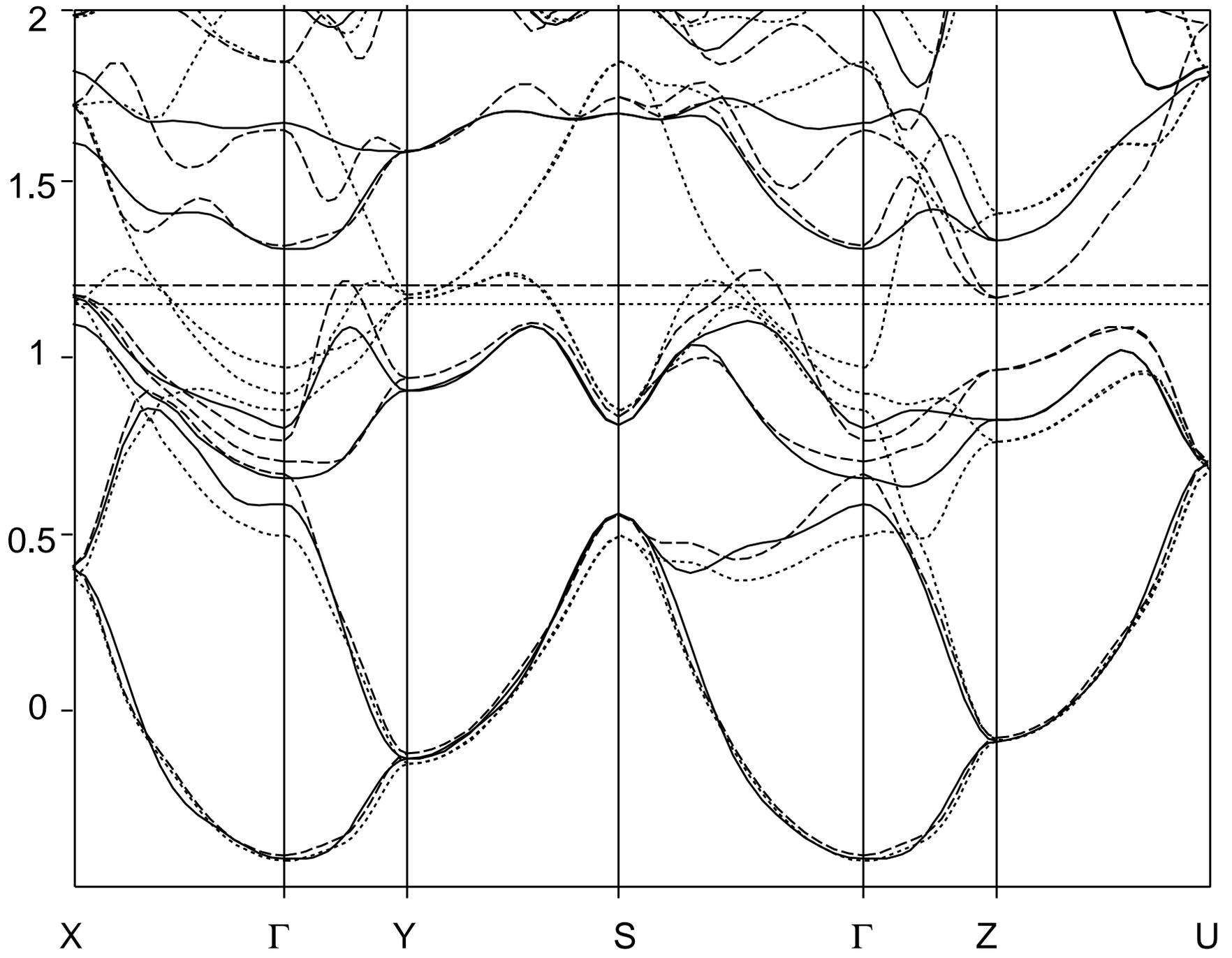
\end{document}